\documentclass[twocolumn,aps,physrev,superscriptaddress,nobalancelastpage,citeautoscript,amsmath,amssymb,floatfix]{revtex4-2}

\usepackage{graphicx}
\usepackage{natbib}
\usepackage{xcolor}
\usepackage[export]{adjustbox}
\usepackage{dcolumn}
\usepackage{ulem}
\usepackage{bm}
\usepackage[T1]{fontenc}
\usepackage{multirow}
\usepackage[%
  colorlinks=true,
  urlcolor=blue,
  linkcolor=blue,
  citecolor=blue
]{hyperref}

\begin{document}

\title{Current-induced magnetization switching of exchange-biased NiO heterostructures characterized by spin-orbit torque}
\author{Krzysztof~Grochot}
\email{grochot@agh.edu.pl}
\affiliation{AGH University of Science and Technology, Department of Electronics, al. Mickiewicza 30, 30-059 Krak\'{o}w, Poland}
\affiliation{AGH University of Science and Technology, Faculty of Physics and Applied Computer Science, al. Mickiewicza 30, 30-059 Kraków, Poland}
\author{Łukasz Karwacki}
\email{karwacki@ifmpan.poznan.pl}
\affiliation{ Institute of Molecular Physics, Polish Academy of Sciences, ul. Smoluchowskiego 17, 60-179 Pozna\'{n}, Poland}
\author{Stanisław Łazarski}
\affiliation{AGH University of Science and Technology, Department of Electronics, al. Mickiewicza 30, 30-059 Krak\'{o}w, Poland}
\author{Witold Skowroński}
\affiliation{AGH University of Science and Technology, Department of Electronics, al. Mickiewicza 30, 30-059 Krak\'{o}w, Poland}
\author{Jarosław Kanak}
\affiliation{AGH University of Science and Technology, Department of Electronics, al. Mickiewicza 30, 30-059 Krak\'{o}w, Poland}
\author{Wiesław Powroźnik}
\affiliation{AGH University of Science and Technology, Department of Electronics, al. Mickiewicza 30, 30-059 Krak\'{o}w, Poland}
\author{Piotr Kuświk}
\affiliation{ Institute of Molecular Physics, Polish Academy of Sciences, ul. Smoluchowskiego 17, 60-179 Pozna\'{n}, Poland}
\author{Mateusz Kowacz}
\affiliation{ Institute of Molecular Physics, Polish Academy of Sciences, ul. Smoluchowskiego 17, 60-179 Pozna\'{n}, Poland}
\author{Feliks Stobiecki}
\affiliation{ Institute of Molecular Physics, Polish Academy of Sciences, ul. Smoluchowskiego 17, 60-179 Pozna\'{n}, Poland}
\author{Tomasz Stobiecki}
\affiliation{AGH University of Science and Technology, Department of Electronics, al. Mickiewicza 30, 30-059 Krak\'{o}w, Poland}
\affiliation{AGH University of Science and Technology, Faculty of Physics and Applied Computer Science, al. Mickiewicza 30, 30-059 Kraków, Poland}

\date{\today}
\begin{abstract}
In this work, we study magnetization switching induced by spin-orbit torque in W(Pt)/Co/NiO heterostructures with variable thickness of heavy-metal layers W and Pt, perpendicularly magnetized Co layer and an antiferromagnetic NiO layer. Using current-driven switching, magnetoresistance and anomalous Hall effect measurements, perpendicular and in-plane exchange bias field were determined. Several Hall-bar devices possessing in-plane exchange bias from both systems were selected and analyzed in relation to our analytical switching model of critical current density as a function of Pt and W thickness, resulting in estimation of effective spin Hall angle and perpendicular effective magnetic anisotropy. We demonstrate in both the Pt/Co/NiO and the W/Co/NiO systems the deterministic Co magnetization switching without external magnetic field which was replaced by in-plane exchange bias field. Moreover, we show that due to a higher effective spin Hall angle in W than in Pt-systems the relative difference between the resistance states in the magnetization current switching to difference  between the resistance states in magnetic field switching determined by anomalous Hall effect ($\Delta R/\Delta R_{\text{AHE}}$) is about twice higher in W than Pt, while critical switching current density in W is one order lower than in Pt-devices. The current switching stability and training process is discussed in detail.
\end{abstract}

\maketitle

\section{Introduction}
\label{sec:intro}
Spin-orbit torque random access memories (SOT-RAMs) are anticipated as a next-generation of low-power, high-endurance, non-volatile and energy-efficient magnetic RAMs, which fit into the modern trend of green information technology (IT)~\cite{brataas_spinorbit_2014, wang_spinorbit_2017}.
Spintronic data storage devices, in contrast to their conventional semiconductor counterparts, need not to be continuously refreshed, leading to the reduction of heat dissipation and lower energy consumption \cite{dieny_opportunities_2020}.  
Recently, SOT-based technologies have evolved as one of the most promising, because they require neither high current densities nor high voltages applied to the thin tunnel barriers~\cite{garello_ultrafast_2014, fukami_spinorbit_2016}, and enable magnetization switching below 1 ns \cite{grimaldi_single-shot_2020}. Such memory cells constitute an efficient alternative to STT-MRAM.

A significant progress has been achieved in understanding and utilizing spin Hall effect (SHE)~\cite{hirsch_spin_1999, zhang_spin_2000,Sinova_2015} in heavy metals (HMs) or topological insulators~\cite{fan_magnetization_2014} to control magnetic states of ferromagnets (FMs) and antiferromagnets (AFMs)~\cite{manchon_current-induced_2019}. The mechanism relies on SOT-induced switching due to accumulated spin density noncollinear with magnetization. However, the torque itself cannot switch the magnetization between two stable states without the up-down degeneracy along the charge current flow direction being broken. It can be achieved by applying an external magnetic field collinear with the current (but noncollinear with the magnetization), which, however, is impractical in device applications and technologically unattractive. Several approaches have been proposed to replace external magnetic field and achieve field-free switching: 
magnetization switching controlled by electric field in hybrid ferromagnetic/ferroelectric structure \cite{cai_electric_2017}, two coupled FM layers exhibiting magnetization easy axes orthogonal to each other~\cite{lau_spinorbit_2016,lazarski_field-free_2019,baek_spin_2018,sheng_adjustable_2018,  wang_field_2018,  chuang_cr_2019, cao_tuning_2019} or introducing a lateral symmetry breaking by asymmetric layers~\cite{chen_free_2019, safeer_spinorbit_2016, you_switching_2015, yu_current-driven_2014, yu_switching_2014}, among others.
However, one of the most promising solutions is still the well-known exchange bias induced by interfacial exchange coupling a ferromagnet with an antiferromagnetic layer~\cite{maat_perpendicular_2001, oh_field-free_2016, fukami_magnetization_2016, van_den_brink_field-free_2016}. The use of metallic antiferromagnet for this purpose has already been described in the literature~\cite{razavi_joule_2017, lau_spinorbit_2016} and it was shown to support both the spin Hall effect and exchange bias in a single layer. This setup, however, makes further optimization of SOT-switching of ferromagnet difficult as the electron spin density generated in AFM acts not only on the ferromagnet but on N\'eel order as well \cite{manchon_spin_2017}. To distinguish between torques acting on ferromagnetic layers from the spin-orbit--induced effect and the exchange bias effect, one can use a heavy-metal/ferromagnet coupled with antiferromagnetic insulator such as NiO. It not only induces exchange bias~\cite{kuswik_prb_2018, kuswik_tailoring_2018, mazalski_demagnetization_2020} but also can enhance perpendicular magnetic anisotropy (PMA) of the ferromagnetic layer ~\cite{kuswik_enhancement_2016} and allow to achieve lower critical switching current than in the case of metallic AFMs. 

Using HM/FM/NiO heterostructures as one of the elements in M-RAMs is possible, as a recent study on NiO/MgO tunnel junctions has shown that there is a sizeable tunneling magnetoresistance (TMR) \cite{yang_negative_2011}, although its appearance is more complex than using MgO barrier alone - the insertion of NiO leads to the appearance of a strong asymmetry in TMR and in particular to negative TMR \cite{yang_negative_2011}. It has been also shown, that the NiO alone can support large TMR in various MTJs \cite{ono_magnetoresistance_1996, sokolov_resonant_2003}. Moreover, there is a possibility for a novel type of memory cells, as it has been shown that the NiO can mediate anti-damping spin-transfer torque between metallic layers \cite{moriyama_anti-damping_2015}.

Motivated by above considerations we present here the study of magnetization switching induced by spin-orbit torque in W(Pt)/Co/NiO heterostructures with variable thickness of W and Pt layers, perpendicularly magnetized Co layer and an antiferromagnetic NiO layer.
Using magnetoresistance measurements and current-driven magnetization switching, we demonstrate simultaneous occurrence of in-plane ($H_{\text{exb}}^{\text{(x)}}$) and perpendicular ($H_{\text{exb}}^{\text{(z)}}$) components of exchange bias field. We show Co magnetization switching without external magnetic field which was replaced by in-plane exchange bias field and for this case we developed analytical magnetization switching model of critical current density. Finally, we discuss the current switching stability and training process carried out on Hall-bar devices of Pt/Co/NiO and W/Co/NiO. 

The remainder of this paper is organized as follows: Section II provides details of sample fabrication, and explains the experimental techniques used to characterize the samples, Sec. III describes theoretical model for spin Hall threshold currents adopted for exchange biased samples. Section IV contains the results and their discussions. Finally, Sec. VI concludes and summarizes the paper.

\section{Experiment}
\label{sec:exp}
\begin{figure}[h]
    \centering
	\includegraphics[width=1\columnwidth]{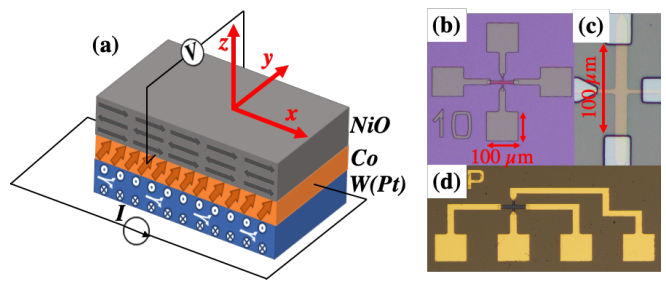}
	\caption{(a) Schematic representation of our multilayer system. The orange arrows indicate perpendicularly magnetized Co layer. The out-of-plane vectors on W(Pt)/Co interfaces show the accumulated spin as a result of the spin Hall effect. Optical microscopic image of the patterned Hall-bar device: (b) Hall-bar for magnetoresistance measurements, (c) detailed dimensions of the Hall-bar, (d) Hall-bar for SOT-induced magnetization switching measurements. }
	\label{fig:elements}
\end{figure}

Two HM/FM/AFM multilayer systems consisting of two different heavy metals: W and Pt were deposited. As shown schematically in Fig.~\ref{fig:elements}(a) the bottom-up heterostructure is sequenced as: Si/$\mathrm{SiO_2}$/W(Pt)/Co/NiO. Heavy-metal layer was deposited in wedge shaped form with thickness varied from 0 to 10 nm along 20 mm long sample edge (x coordinate). The resulting thickness gradient was achieved by controlled movement of a shutter. The thicknesses of the two other layers namely Co and NiO were 0.7 nm and 7 nm, respectively. We also deposited Pt(4)/Co(1)/MgO system (thicknesses in nm) as a reference sample for further analysis. All metallic layers were deposited by magnetron sputtering at room temperature. 

In the case of W sputtering, low DC power of 4 W and 6 cm target-sample distance were used, which resulted in a deposition rate of 0.01 nm/s. Such conditions are essential for the growth of W layer in the cubic $\mathrm{\beta}$ phase. The Pt and Co were deposited with DC power of 8 and 15 W, respectively. 

The stoichiometric NiO layer deposited on the top of Co layer was prepared from NiO target using a pulsed laser deposition technique (PLD). The process was performed in a controlled oxygen atmosphere under $\mathrm{O_2}$ partial pressure $\mathrm{1.5 \cdot 10^{-5}}$ mbar in a separate UHV chamber and samples were transferred between chambers without breaking UHV conditions. Systematic studies of the perpendicular exchange bias effect in Au/Co/NiO/Au system by magnetooptical Kerr rotation, published in the work \cite{kuswik_enhancement_2016}, have shown that Co layer is oxidized due to deposition of NiO in an oxygen-rich atmosphere. Our X-ray absorption spectroscopy studies (for details see Supplemental Materials \cite{supplemental}) also confirm surface oxidation Co layer to stoichiometric CoO. The oxidation process effectively reduces the ferromagnetic thickness of Co which leads to an increase in the surface anisotropy finally revealing in strong PMA of Co. Moreover, interlayer exchange bias coupling appeared when perpendicular magnetic field of 1.1 kOe was applied during deposition. 

Thicknesses of all layers were determined from the deposition growth rate of particular materials calibrated using x-ray reflectivity measurements. Next, all as-deposited samples were characterized before patterning by X-ray diffraction $\mathrm{\theta - 2 \theta}$ (XRD) and grazing incidence diffraction (GIXD) (for details see Supplemental Materials~\cite{supplemental}). All systems were also examined by the polar Kerr magnetometer (p-MOKE) to determine the range of HM thicknesses for which PMA occurs. Square-shape hysteresis loops were observed, which indicate the presence of PMA in both systems for Pt layer thickness $t_{\text{Pt}}$ between 1 and 9 nm and for W layer thickness $t_{\text{W}}$ between 3.5 and 8 nm, which was next confirmed by anomalous Hall effect (AHE) measurements (see in Supplemental Materials~\cite{supplemental}).

After basic characterization of continuous samples, both heterostructures were patterned using optical direct imaging lithography (DLP) and ion etching to create a matrix of Hall-bar devices, with different $t_{\text{HM}}$ for subsequent electrical measurements (Fig.~\ref{fig:elements}(b-d)). The sizes of prepared structures were 100 $\mathrm{\mu m}$ $\times$ 10 $\mathrm{\mu m}$ for magnetoresistance and AHE measurements and 30 $\mathrm{\mu m}$ $\times$ 30  $\mathrm{\mu m}$ for current-induced magnetization switching experiments. Al(20)/Au(30) electrical leads of 100 $\mathrm{\mu m}$ $\times$ 100 $\mathrm{\mu m}$ were deposited in a second lithography step followed by the lift-off process. Specific locations of pads near the Hall-bars were designed for measurement in custom-made rotating probe station allowing 2- or 4-points measurement of electrical transport properties in the presence of the magnetic field applied at arbitrary azimuthal and polar angle with respect to the Hall-bar axis.

The resistance of each Hall-bar was measured using a four-point method~\cite{smits1958} and resistivities of Pt and W layers were determined using a parallel resistors model and the method described by Kawaguchi et al.~\cite{kawaguchi_2018}. The Pt and W resistivities analysis yielded 30 $\mathrm{\mu \Omega cm}$~\cite{kawaguchi_2018, Sagasta_2016,lazarski_field-free_2019, skowronski_2019} and 174 $\mathrm{\mu \Omega cm}$~\cite{hao_beta_2015, hao_giant_2015, neumann2016, skowronski_2019}, respectively. Depending on HM underlayer Co resistivity was 28 $\mathrm{\mu \Omega cm}$ when deposited on Pt~\cite{lazarski_field-free_2019} and 58 $\mathrm{\mu \Omega cm}$ on W. The details of the resistivity measurements are presented in Supplemental Materials~\cite{supplemental}.

\section{Critical current model}
\label{sec:theory}

In order to determine the influence of exchange bias on threshold current in our system we follow the analysis for spin Hall threshold currents first derived by Lee \textit{et al.}~\cite{Lee2013, taniguchi_theoretical_2019}.

We start with Landau-Lifshitz-Gilbert (LLG) equation for macrospin  magnetization $\hat{\mathbf{m}}=(m_x,m_y,m_z)=(\cos \phi \sin \theta, \sin \phi \sin \theta, \cos \theta)$,
\begin{equation}
    \label{eq:LLG}
\frac{d \hat{\mathbf{m}}}{dt}-\alpha \hat{\mathbf{m}} \times \frac{d \hat{\mathbf{m}}}{d t}=\mathbf{\Gamma}\,,
\end{equation}
where $\alpha$ is Gilbert damping constant.

The general torque exerted on magnetization assumes the following form
\begin{equation}
    \label{eq:torque}
\mathbf{\Gamma}=-\gamma_{0} \hat{\mathbf{m}} \times \mathbf{H}_{\mathrm{eff}}-\gamma_{0} H_{\mathrm{DL}} \hat{\mathbf{m}} \times \hat{\mathbf{m}} \times \hat{\mathbf{y}}\,,
\end{equation}
where $\gamma_0$ is the gyromagnetic constant, and the first term comes from the effective field, $\bm{H}_{\text{eff}}=-\nabla_{\bm{m}}u$,

where free energy of the FM has the following form
\begin{equation}
    \label{eq:energy}
u = - \frac{1}{2} H_{\text{K,eff}} m_z^2 - \frac{1}{2} H_\text{A} m_y^2 - m_xH_{x} - m_x H_{\text{exb}}^{(x)}\,,
\end{equation}
with $H_{\text{K,eff}}$ being the field of effective perpendicular magnetic anisotropy, $H_\text{A}$ field of effective in-plane anisotropy, $H_{x}$ the magnetic field along $x$, and $H_{\text{exb}}^{(x)}$ the in-plane exchange bias.

The second torque term in Eq.~(\ref{eq:torque}) comes from damping-like field,
\begin{equation}
    \label{eq:dlfield}
H_{\mathrm{DL}} =\frac{\hbar}{2 e \mu_{0} M_{s} t_{\text{FM}}} \theta_{\mathrm{SH}} j_{\text{HM}}\left(1-\operatorname{sech} \frac{t_{\text{HM}}}{\lambda_{\text{HM}}}\right) \frac{g_{r}}{1+g_{r}} \,,
\end{equation}
where $\hbar$ is reduced Planck's constant, $e$ is the elementary charge, $\mu_0M_s$ is magnetization, $t_{\text{FM}}$ is thickness of ferromagnetic layer, $\theta_{\text{SH}}$ the spin Hall angle, $j_{\text{HM}}$ the current density flowing through HM, and $g_{r}=2\lambda_{\text{HM}} \rho_{\text{HM}} G_{r} \operatorname{coth} \left(t_{\text{HM}}/\lambda_{\text{HM}}\right)$ is the unitless real part of spin-mixing conductivity, $G_r$ with $\lambda_{\text{HM}}$, $\rho_{\text{HM}}$, and $t_{\text{HM}}$ the HM's spin diffusion length, resistivity, and thickness, respectively. The damping-like field is induced by $\hat{\mathbf{y}}$-polarized spin accumulation due to spin Hall effect in HM.

Stationary solution of the LLG equation (\ref{eq:LLG}) leads to the torque equilibrium condition, $\mathbf{\Gamma}=0$.
For strong magnetic field applied along $x$ we assume $\phi\approx 0$ which leads to the following condition for damping-like field
\begin{equation}
H_{\mathrm{DL}}=\cos \theta\left(H_{\text{exb}}^{(x)}-H_{K, \mathrm{eff}} \sin \theta+H_{x}\right)\,.
\end{equation}
By analyzing stability of the above equation we obtain simplified relation for critical damping-like field
\begin{align}
H_{\mathrm{DL}}^{\mathrm{sw}} \approx \frac{H_{K, \mathrm{eff}}}{2}-\frac{H_{x}-H_{\text{exb}}^{(x)}}{\sqrt{2}}\,.
\end{align}
Inserting into the above equation explicit formula for damping like field, Eq.~(\ref{eq:dlfield}), we obtain for critical current density the following expression
\begin{align}
j_{c}^{\mathrm{sw}} \approx \frac{2 e \mu_{0} M_{s} t_{\text{FM}}\left(1+g_{r}\right)}{\hbar \theta_{\mathrm{SH}} g_{r}\left(1-\operatorname{sech} \frac{t_{\text{HM}}}{\lambda_{\text{HM}}}\right)}\left(\frac{H_{K, \mathrm{eff}}}{2}-\frac{H_{x}-H_{\text{exb}}^{(x)}}{\sqrt{2}}\right)\,.
\label{eq:current_density_raw}
\end{align}
Assuming perfect HM/FM interface, i.e. $G_r\rightarrow\infty$, and assuming $t_{\text{HM}}\gg\lambda_{\text{HM}}$ leads to the simplified expression,
\begin{align}
j_{c}^{\mathrm{sw}} \approx \frac{2 e \mu_{0} M_{s} t_{\text{FM}}}{\hbar \theta_{\mathrm{SH}}}\left(\frac{H_{K, \mathrm{eff}}}{2}-\frac{H_{x}-H_{\text{exb}}^{(x)}}{\sqrt{2}}\right)\,,
\label{eq:current_density}
\end{align}
which is used later on to fit the experimental data. Note, that our model does not take into account switching mechanism due to creation and motion of domain walls, which has been observed in Pt/Co system \cite{baumgartner_spatially_2017} and results in smaller switching current density than the one estimated by the above model. Our estimate can, however, be treated as the upper limit.

\section{Results and discussion}
\label{sec:results}

\subsection{SOT Current Switching}

Anomalous Hall effect was used to determine the current-driven magnetization switching between high and low stable resistance states. The measurement setup is shown in Fig.~\ref{fig:elements}(a). Initially, the sample has been magnetized by external magnetic field applied along \textit{z} direction to the state corresponding to low resistance of the AHE loop.
\begin{figure}[h!]
	\centering
	\includegraphics[width=1\linewidth]{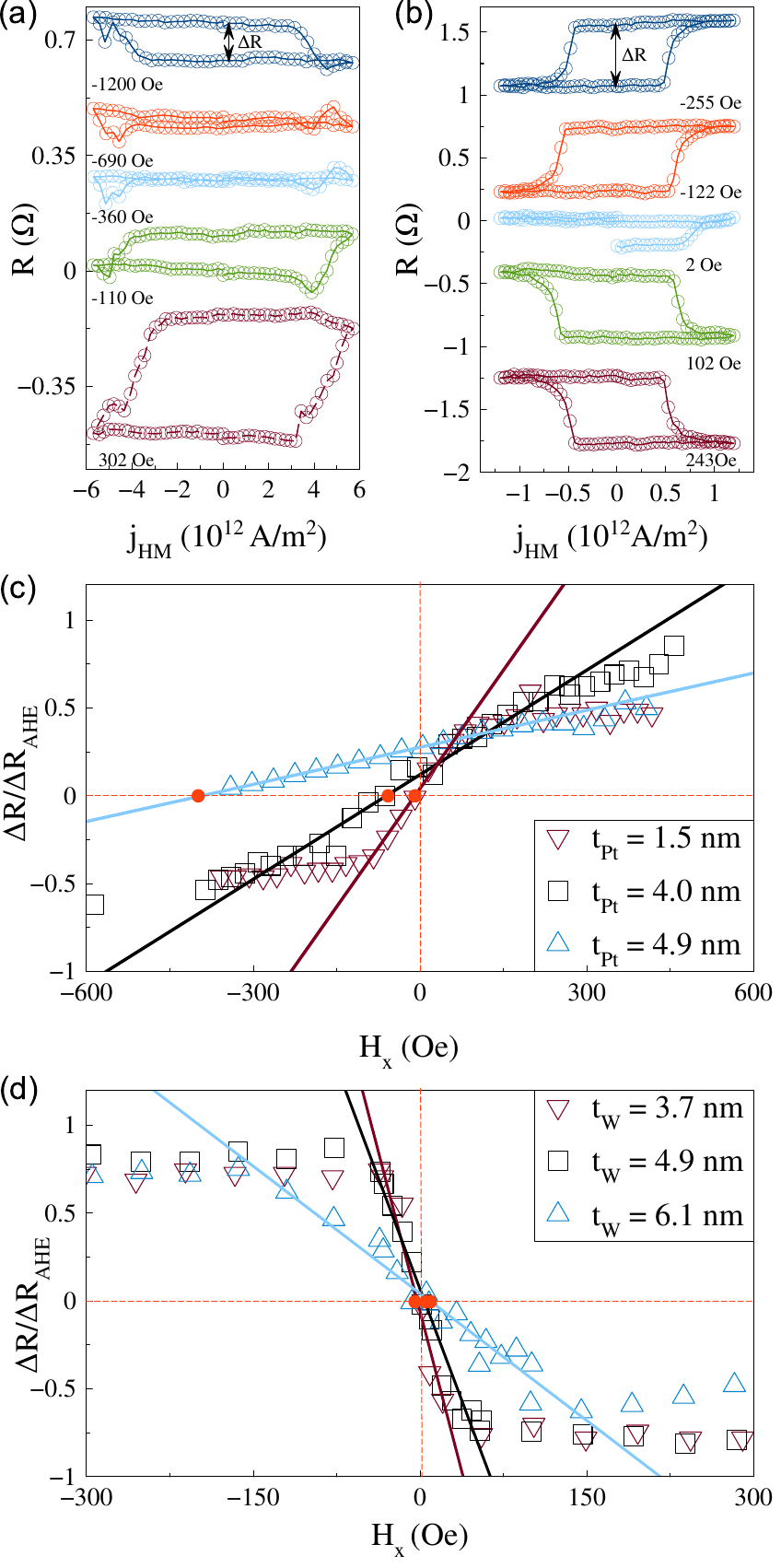}
	\caption{Examples of the current switching loops for different values of external magnetic field $H_{\text{x}}$ in Pt(4.9)/Co/NiO (a) and W(4.9)/Co/NiO (b) system. The ratio $\Delta R/R_{AHE}$ are depicted in (c) and (d), the intersection is marked by a red dot. $H_{\text{exb}}^{(x)}$ in Pt system is 5 Oe for $t_{\text{Pt}} = 1.5~\text{nm}$, 60 Oe for $t_{\text{Pt}} = 4.0~\text{nm}$ and 392 Oe for $t_{\text{Pt}} = 4.9~\text{nm}$, respectively. For all analyzed W elements these values are approximately 0 Oe in the measurement error limit. }
	\label{fig:loops}
\end{figure}

Then, a sequence of current pulses with 10 ms duration and 20 ms intervals in the \textit{x} direction was applied to drive the magnetization switching. The current was swept from negative to positive and back to negative and simultaneously transversal voltage was measured in presence of in-plane magnetic field, collinear with the current direction ($H_{\text{x}})$. The value of $H_{\text{x}}$ was changed sequentially after each switching loop. 

As a result, we obtained the current switching loops for Pt- and W-based Hall-bar devices (Fig.~\ref{fig:loops}(a),(b)). Opposite loops polarities result that Pt has a positive spin Hall angle and W a negative one.
\begin{figure}[h!]
 	\centering
 	\includegraphics[width=1\columnwidth]{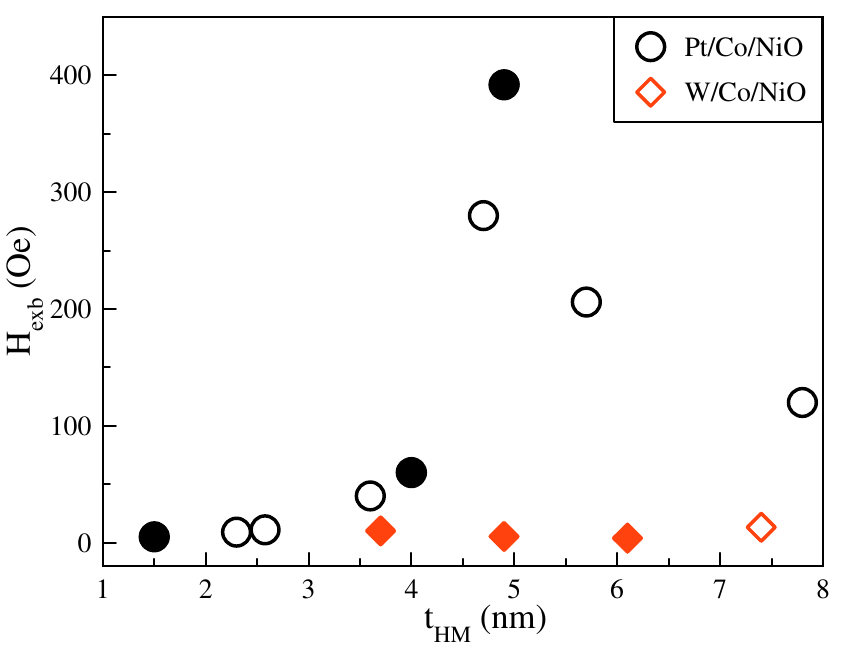}
 	\caption{$H_{\text{exb}}^{(x)}$ obtained from $\Delta R$ zero shifts for as-deposited W(Pt)/Co/NiO system. Filled points mark the elements for which the theoretical threshold current model was fitted. }
 	\label{fig:exb_in_plane}
 \end{figure}
 
By analyzing the difference between high and low AHE resistance from AHE loop for different thickness of Pt and W-based devices $\Delta R_{\text{AHE}}$ was determined. Next, the current switching loop opening $\Delta R$ was measured, for each value of applied magnetic field $H_{\text{x}}$, which shows closing  loop with decreasing $H_{\text{x}}$ (Fig.~\ref{fig:loops}(a),(b)). The $\mathrm{\Delta R / \Delta R_{AHE}}$ ratio as a function of $H_{\text{x}}$ which is a measure of the effectiveness of current induced magnetization switching for Pt and W-devices is shown in Fig.~\ref{fig:loops}(c),(d). Obtained dependence is approximately linear for small positive and negative $H_{\text{x}}$. As can be seen, $\mathrm{\Delta R}$ does not reach $\mathrm{\Delta R_{AHE}}$, even when large magnetic field is applied (See also Fig. S10 in Supplemental Materials \cite{supplemental}).

The intersection of linear function with zero $\mathrm{\Delta R / \Delta R_{AHE}}$, corresponding to the magnetic field  for which the loop is closed, can be identified as the value of $H_{\text{x}}$ which compensates the in-plane component of exchange bias field ($H_{\text{exb}}^{(x)}$) and allows to indirectly determine the value of $H_{\text{exb}}^{(x)}$, because the mentioned in-plane compensation field has the same value, but the opposite sign.

The intersection points for Pt-based system depend on Pt layer thickness, reaching maximal compensation field magnitude for $t_{\text{Pt}}$=4.9 nm, while in W-based structure intersection occurs roughly at zero $H_{\text{x}}$ field in a wide range of W layer thicknesses as shown in Fig.~\ref{fig:exb_in_plane}. In high $H_{\text{x}}$ fields the $\Delta R$ saturates reaching for W-devices about $0.8\Delta R_{\text{AHE}}$ (Fig.~\ref{fig:loops}(d)), while for Pt-devices it changes with increasing Pt thickness from about $0.5\Delta R_{\text{AHE}}$ to $0.28\Delta R_{\text{AHE}}$ (Fig.~\ref{fig:loops}(c) and Fig. S10(a) in Supplemental Materials \cite{supplemental}). 
 The perpendicular exchange bias field ($H_{\text{exb}}^{(z)}$), determined by AHE hysteresis is highest for about 5 nm of HM thickness in both systems, but $H_{\text{exb}}^{(z)}$ in W-based system is two times smaller than in Pt-based system (see Fig. S7 in Supplemental Materials \cite{supplemental}).

Because in as-deposited W/Co/NiO heterostructure the magnitude of $H_{\text{exb}}^{(x)}$ is negligible, field-free SOT magnetization switching is not achieved. Therefore, to induce the in-plane component of exchange bias, the system was annealed at 100 $\mathrm{^oC}$ (i.e. at a temperature slightly higher than blocking temperature of 373K but lower than N\'eel temperature of 525 K) for 15 min and then cooled to room temperature in the presence of external magnetic field of 4 kOe applied perpendicularly to the sample. Afterwards, the re-measured AHE loop for selected HM thickness indicates the presence of the PMA and $H_{\text{exb}}^{(z)}$ in Co layer, manifested by rectangular shape shifted of -67 Oe (see Fig. S6(b) in Supplemental Materials \cite{supplemental}). In the next step the current switching experiments were repeated and results were analyzed as described above. Finally, $H_{\text{exb}}^{(x)}$ = -148 Oe was obtained from $\Delta R$ measurements (see Fig. S10(b) in Supplemental Materials \cite{supplemental}). For further analysis and fitting our threshold current model in exchange-biased system, three as-deposited Pt/Co/NiO Hall-bar devices of 1.5, 4.0 and 4.9 nm Pt thicknesses were selected and identified as A2, A3 and A4, respectively. We selected also three as-deposited W/Co/NiO Hall-bars of 3.7, 4.9 and 6.1 nm W thicknesses which were marked as B1, B2 and B3, respectively, and chose annealed W(4.3)/Co/NiO device (C1). We also fitted our model to reference sample (labeled A1) which was used to verify the model, as indicated earlier.
 
\begin{table*}[t]
	\caption{The table shows values of in-plane exchange bias field ($H_{\text{exb}}^{(x)}$), effective anisotropy field ($H_{\text{K,eff}}$), spin Hall angle $\theta_{\text{SH,eff}}$ and saturation magnetization ($\mu_{\text{0}} M_{\text{s}}$) obtained as a result of magnetoresistance measurements and fitting the threshold current model to the experimental data.  }

	\begin{ruledtabular}
  
\begin{tabular}{cccccc}
\multicolumn{6}{c}{Pt/Co/MgO}                                                                                                                                                                                                                                         \\ 
\hline
\multicolumn{1}{l}{Sample} & \multicolumn{1}{l}{$t_{\text{Pt}}$~(nm)~} & \multicolumn{1}{l}{$H_{\text{exb}}^{(x)}$~(Oe)~} & \multicolumn{1}{l}{$H_{\text{K,eff}}$—FIT (Oe)~} & \multicolumn{1}{l}{$\theta_{\text{SH,eff}}$—FIT (\%)} & \multicolumn{1}{l}{$\mu_0 M_s$~(T)~}  \\ 
\hline
A1                         & 4.0                                       & 0                                         & 2508$\pm$~80                                     &13.5~$\pm$~1                                          & 0.5                                   \\
\hline
\multicolumn{6}{c}{Pt/Co/NiO}                                                                                                                                                                                                                                         \\ 
\hline
\multicolumn{1}{l}{Sample} & \multicolumn{1}{l}{$t_{\text{Pt}}$ (nm) } & \multicolumn{1}{l}{$H_{\text{exb}}^{(x)}$ (Oe) } & \multicolumn{1}{l}{$H_{\text{K,eff}}$—FIT (Oe) } & \multicolumn{1}{l}{$\theta_{\text{SH,eff}}$—FIT (\%) } & \multicolumn{1}{l}{$\mu_0 M_s$ (T) }  \\ 
\hline
A2                         & 1.5                                       & 6 $\pm$ 1                                 & 2141 $\pm$ 10                                    & 4.1 $\pm$ 0.8                                         & 0.5                                   \\
A3                         & 4.0                                       & 176 $\pm$ 3                               & 4130 $\pm$ 10                                    & 5.2 $\pm$ 1.2                                         & 0.5                                   \\
A4                         & 4.9                                       & 522 $\pm$ 14                              & 4638 $\pm$ 10                                    & 5.8 $\pm$ 1.3                                         & 0.5                                   \\ 
\hline
\multicolumn{6}{c}{W/Co/NiO}                                                                                                                                                                                                                                          \\ 
\hline
\multicolumn{1}{l}{Sample} & \multicolumn{1}{l}{$t_{\text{W}}$ (nm) }  & \multicolumn{1}{l}{$H_{\text{exb}}^{(x)}$ (Oe) } & \multicolumn{1}{l}{$H_{\text{K,eff}}$—FIT (Oe) } & \multicolumn{1}{l}{$\theta_{\text{SH,eff}}$—FIT (\%) } & \multicolumn{1}{l}{$\mu_0 M_s$ (T) }  \\ 
\hline

B1                         & 3.7                                       & 0                                         & 1132 $\pm$ 20                                    & -5.7 $\pm$ 1.1                                        & 0.5                                   \\
B2                         & 4.9                                       & 30 $\pm$ 15                               & 1031 $\pm$ 10                                    & -7.5 $\pm$ 1.5                                        & 0.5                                   \\
B3                         & 6.1                                       & 7 $\pm$ 1                                 & 1035 $\pm$ 11                                    & -9.3 $\pm$ 1.8                                        & 0.5                                  \\
C1                         & 4.3                                       & -158 $\pm$ 29                             & 2584 $\pm$ 2                                     & -44.0 $\pm$ 5                                         & 0.5                                   \\
\end{tabular}

\end{ruledtabular}
\label{tab:fit}
\end{table*}

\subsection{In-plane exchange bias}
\label{sec:bias}

In order to confirm the above-discussed $H_{\text{exb}}^{(x)}$, resistance along the Hall-bar ($R_{xx}$) was measured with external magnetic field being swept along the $x$ direction and modeled it with the equation:
\begin{equation}
    \label{eq:rxx}
R_{xx} = R_0 + \Delta R_{\text{AMR}} m_x^2\,,
\end{equation}
where $R_0$ is magnetization-independent resistance, and $\Delta R_{\text{AMR}}$ denotes changes due to anisotropic magnetoresistance effect.
Considering equilibrium condition of energy density (Eq.\ref{eq:energy}) with respect to angle $\phi$ longitudinal resistance was reformulated to:
\begin{equation}
    \label{eq:rxx_h}
R_{xx} \approx R_0 + \Delta R_{\text{AMR}}  \frac{(H_{\text{exb}}^{(x)}+H_x)^2}{H_\text{A}^2}\,.
\end{equation}

\begin{figure}[h!]
	\centering
	\includegraphics[width=1 \linewidth]{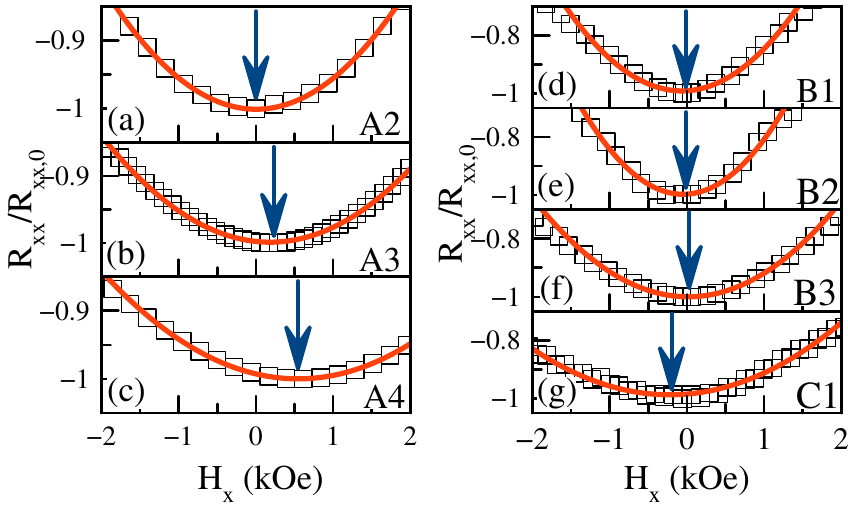}
	\caption{$H_{\text{exb}}^{(x)}$ obtained from $R_{\text{xx}}(H_{\text{x}})$ measurements for as-deposited Pt- (a-c) and W-based (d-f) systems and annealed  W/Co/NiO system (g). The blue arrows indicate the position of in-plane $H_{\text{exb}}^{(x)}$. Experimental data have been normalized to a minimum point. The red lines are fits according to Eq.(10). }
	\label{fig:exb_amr}
\end{figure}
Measured $R_{\text{xx}}$ for A2-A4, B1-B3 and C1 samples are shown in Fig.~\ref{fig:exb_amr}. A parabolic function has been fitted to the data points and minima of the functions are indicated with arrows. According to the Eq.~(\ref{eq:rxx_h}), the minima can be identified as $H_{\text{exb}}^{(x)}$ field, like these discussed in previous section. The resulting values for A2, A3 and A4 samples of about 6 Oe, 176 Oe and 522 Oe (Fig.~\ref{fig:exb_amr}(d)-(f)), respectively, are consistent with the ones obtained in $\Delta R$ opening loop of the current switching experiment described in Sec. IV A. As-deposited B-series Hall-bars still exhibit negligible loop shifts (Fig.~\ref{fig:exb_amr}(d)-(f)). The only exception is the annealed C1 element for which the value is -158 Oe (Fig.~\ref{fig:exb_amr}(g)). $H_{\text{exb}}^{(x)}$ values obtained from magnetoresistance (MR) measurements are in general less noisy and are used for further analysis.
\begin{figure}[h!]
	\centering
	\includegraphics[width=0.8\linewidth]{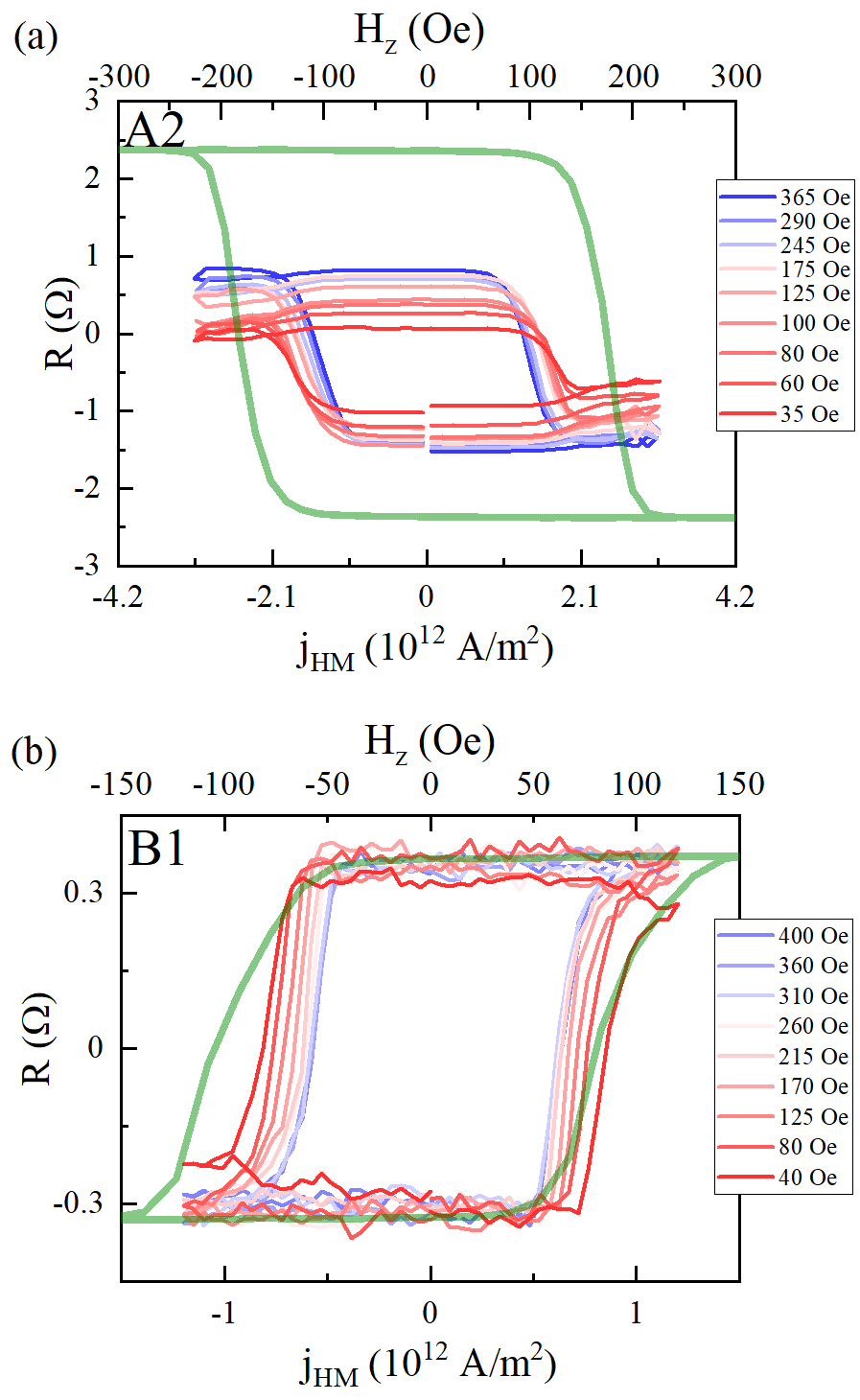}
	\caption{Examples of a series of current switching hysteresis loops for sequentially changed values of $H_{\text{x}}$ in two selected elements A2 (a) and B1 (b), respectively. Anomalous Hall resistances are represented by the green line. }
	\label{fig:raw_loops}
\end{figure}
\subsection{Fitting procedure}

For the analysis of the SOT-induced magnetization switching, we used AHE resistance hysteresis loops vs. applied current densities in HM layer ($j_{\text{HM}}$) measured in a different external magnetic field $H_{\text{x}}$ - examples are depicted in Fig.~\ref{fig:raw_loops}.

\begin{figure*}[ht!]
	\centering
	\includegraphics[width=0.9\linewidth]{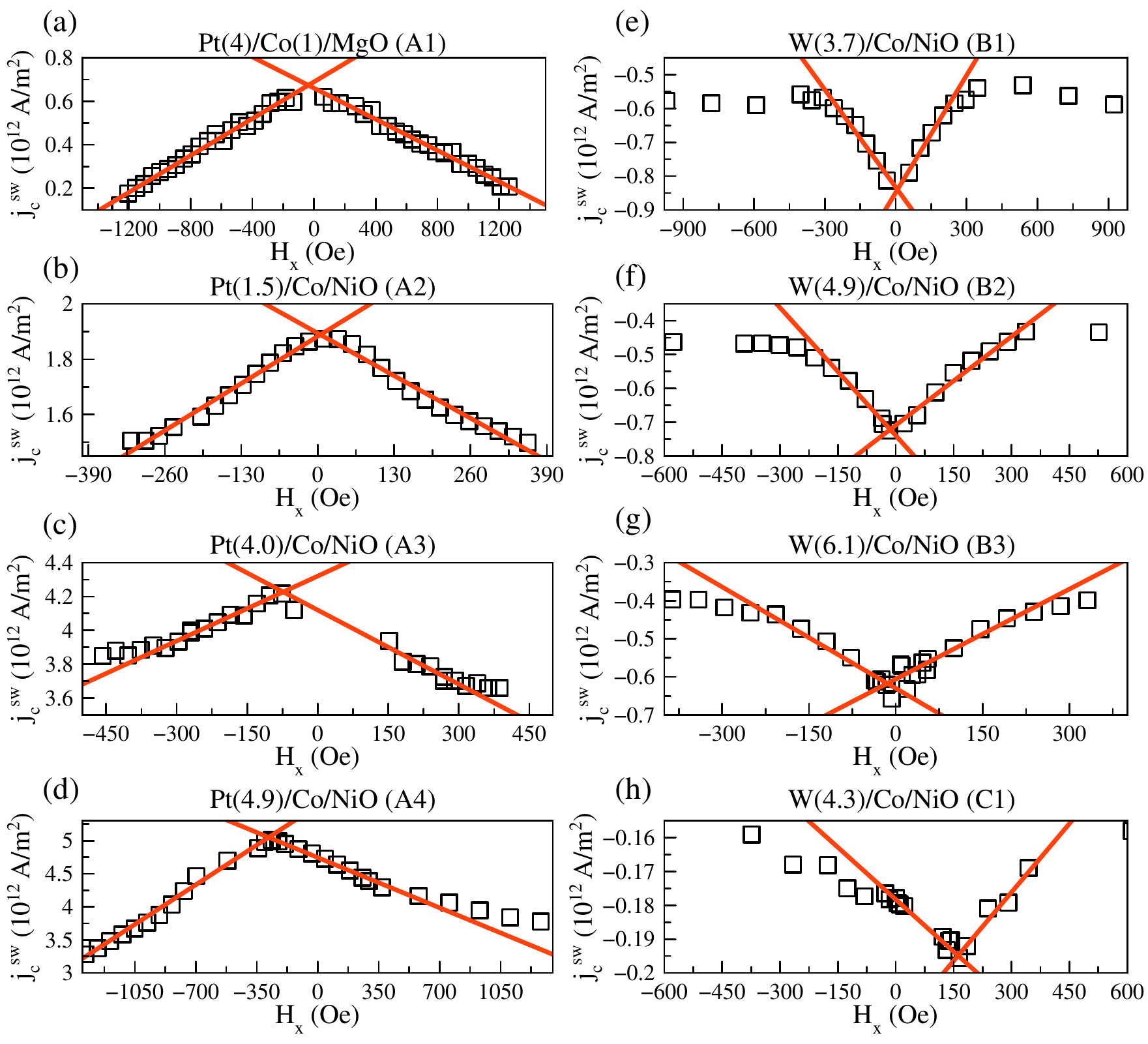}
	\caption{Critical switching current densities, $j_{c}^{\mathrm{sw}}$, as a function of applied external magnetic field $H_{\text{x}}$, for Pt/Co/NiO (a-d) and W/Co/NiO (e-h) differing by HM layer thickness. In the (h) depicted annealed W system. Red line represents model equation fitted to the data points. Parameters used for theoretical model lines are gathered in Tab.\ref{tab:fit}.}
	\label{fig:model_fit}
\end{figure*}
First, using a derivative of $\frac{\partial V_{\text{AHE}}}{ \partial j_{\text{HM}}}$, the threshold switching current ($j_{c}^{\mathrm{sw}}$) was calculated separately for each $H_{\text{x}}$. As a result, linear dependencies of $j_{c}^{\mathrm{sw}}$ vs. $H_{\text{x}}$ are obtained for selected Pt and W thicknesses. Nevertheless, a large number of free parameters in the model equation (Eq. (\ref{eq:current_density})) may cause large uncertainties in the determined values. For this purpose, the first part of Eq. (\ref{eq:current_density}) was replaced by a single parameter $a$ which was fixed and calculated as the linear slope coefficient determined by numerical differentiation of  $j_{c}^{\mathrm{sw}}$ dependence. Therefore, Eq.~(\ref{eq:current_density}) can be rewritten to form:

\begin{equation}
j_c^{sw} \approx a\left(\frac{H_{K,eff}}{2}-\frac{H_x-H_{\text{exb}}^{(x)}}{\sqrt{2}}\right)\,,
\label{eq:upr_crit_curr}
\end{equation}

where: $a$ is fixed parameter obtained from differing $j_{c}^{\mathrm{sw}}$ dependence data points.

This ensures, that the only free fit parameter has remained $H_{\text{K,eff}}$. As mentioned earlier, $H_{\text{exb}}^{(x)}$ has been fixed parameter achieved from the MR measurements.

Initially, the model was verified on the A1 reference sample, which is characterized by zero $H_{\text{exb}}^{(x)}$. In this particular case, both $H_{\text{exb}}^{(x)}$ and $H_{\text{K,eff}}$ parameters have been set as free to check validity of parameters received from fitting. As expected, $H_{\text{exb}}^{(x)}$ = 0 and $H_{\text{K,eff}}$ = 2508 Oe were obtained with a very good compliance level of $R^2$ = 0.93.

Next, a simplified model equation (Eq. (\ref{eq:upr_crit_curr})) was fitted to all selected Pt/Co/NiO and W/Co/NiO. As indicated, $H_{\text{exb}}^{(x)}$ parameter was fixed and set accordingly with Tab.\ref{tab:fit}. Fitting results are depicted in Fig.~\ref{fig:model_fit} where the solid lines correspond to the model equation, respectively for all investigated devices. As shown, the model proved good correlation to the data points. All of the $R^2$ coefficients were above 0.90, ensuring low uncertainty level. For higher $H_{\text{x}}$ there are deviations from linear dependence in both systems.

In case of W-devices, the deviations appear at lower $H_{\text{x}}$ field than in Pt. This can be explained by lower $H_{\text{K,eff}}$ (see Tab.~\ref{tab:fit}) for as-deposited W/Co/NiO than Pt/Co/NiO systems. Devices of A-series were characterized by increase $H_{\text{K,eff}}$ and $H_{\text{c}}$ with increasing Pt thickness (Fig.~\ref{fig:model_fit}(b)-(d)). Additionally, annealing of the B-series devices resulted in doubling $H_{\text{K,eff}}$ value by increasing $H_{\text{c}}$.

Finally, we calculated the effective spin Hall angles ($\theta_{\text{SH,eff}}$) in HM layer. For calculations we used the saturation of magnetization ($\mu_{\text{0}} M_s$) equal to 0.5 T in both systems, obtained from VSM measurements (see Fig. S11 in  Supplemental Materials \cite{supplemental}). We also assumed an infinite value of mixing conduction ($g_r$), which is mostly valid for metallic interfaces \cite{Kim2016}.

Adopted assumptions allowed us to calculate the effective spin Hall angle from the following formula:

\begin{equation}
\theta_{\text{SH,eff}} = \frac{2e \mu_{0} M_s t_F}{\hbar a}~.
\label{eq:spinangle}
\end{equation}

Obtained values of $\mathrm{\theta_{SH,eff}}$ are listed in the Tab.~\ref{tab:fit} and agree with the ones found in the literature \cite{lazarski_field-free_2019, skowronski_2019, rojas_2014, pai_2012, Sagasta_2016, takeuchi_2018, cho_large_2015, hao_giant_2015, zang_2016, skowronski_2016, chen_temperature_2018, Bansal2018,mchugh_impact_2020}.
$\theta_{\text{SH,eff}}$ in as-deposited B-devices are slightly higher than values calculated for A-devices systems, which combined with significantly lower $H_{\text{K,eff}}$ result in approximately one order of magnitude smaller critical switching current densities in this system. It is also worth noting that annealing of the W-based system, apart from induce $H_{\text{exb}}^{(x)}$ also enhance $\mathrm{\theta_{SH, eff}}$ to -44 \% \cite{Bansal2018, skowronski_2016, hao_giant_2015, chen_temperature_2018} and the reduce of $j_{c}^{\mathrm{sw}}$ by the order of magnitude (Fig.~\ref{fig:model_fit}(h)). 
In order to confirm high value of $\mathrm{\theta_{SH, eff}}$, a current switching experiment was performed on another annealed sample from the same series. A similar $\mathrm{\theta_{SH, eff}}$ was obtained (see Fig. S9 in Supplemental materials \cite{supplemental}). We attribute a high value of the effective spin Hall angle of W similarly as in the works \cite{skowronski_2019, hao_giant_2015, Bansal2018, chen_temperature_2018}, to the highly resistive $\beta$-W phase. For example $\theta_{\text{SH,eff}}$ $\approx$ -30\% at RT and over -50\% at 50K in \cite{skowronski_2016}, while in \cite{Bansal2018} it is approximately -44\%. Recently, Oliver L. W. McHugh et al. \cite{mchugh_impact_2020} showed, from first-principles calculations, that interstitials dopants of O and N help to stabilize $\beta$-W grains during film deposition and this process leads to high spin Hall angles.

\subsection{Training effect}

\begin{figure}[h]
	\centering
	\includegraphics[width=1\linewidth]{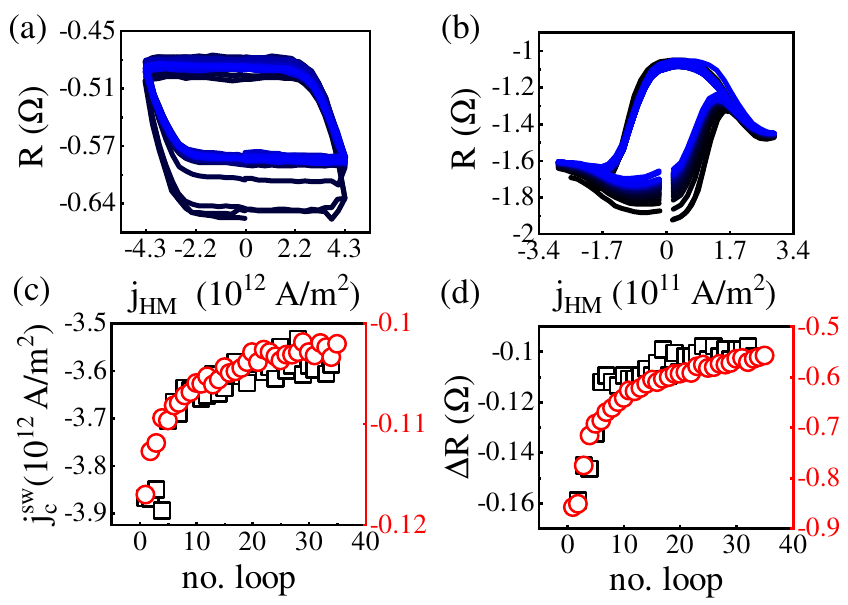}
	\caption{A series of successive switching loops in a measurement sequence for Pt(5.1)/Co(0.7)/NiO (a) and annealed W(4.3)/Co(0.7)/NiO device (b) (thicknesses in nm). The color of each subsequent loop in the sequence changes with increasing magnetic field from black to blue. Critical switching current, $j_{\text{c}}^{\text{sw}}$, vs. number of repeated loops (c) and magnetoresistance, $\Delta R$ (d) for Pt (black points) and W-based devices (red points).}
	\label{fig:stab_loops}
\end{figure}

The training effect in both systems was also investigated for verification of thermal stability of the examined heterostructures. In order to do this 10 times longer pulses than in previous experiments described in Sec. IV A were used. For this purpose, a multiple current switching in a fixed external $H_{\text{x}}$ field was performed by 100 ms current pulses with 200 ms interval between them. The magnitude of the external $H_{\text{x}}$ field has been chosen to obtain an unambiguous magnetization switching. A series of current switching loops depicted in Fig.~\ref{fig:stab_loops}(a),(b) were obtained. Next, loops opening $\Delta R$ and critical current densities $j_{c}^{\mathrm{sw}}$ were determined as a function of loop numbers. The results are presented in Fig.~\ref{fig:stab_loops}(c),(d).
In both systems, $j_{c}^{\mathrm{sw}}$ and $\Delta R$ decrease with increasing number of magnetization switches and their dependence on loop number is similar. During the first few switching events, a significant reduction in both the $\Delta R$ and $j_{c}^{\mathrm{sw}}$ are observed. 
These phenomena can be explained by the progressively increasing temperature in both systems due to Joule heating~\cite{razavi_joule_2017} and training effect~\cite{van_den_brink_field-free_2016, zhang2001, brems2007} witnessed also during magnetic field switching~\cite{binek2004, hochstrat2002, van_den_brink_field-free_2016}.
It is worth mentioning that Joule's heating effect leads to the reduction of the switching current and anisotropy. Saturation of the dependence is caused by achieving a balance between the generated and emitted heating. This saturation occurs for smaller repetition number in the system with Pt (after about 20 switches) in contrast to W-based system in which it is not reached even after 35 switches.

\begin{figure}[h!]
	\centering
	\includegraphics[width=1\linewidth]{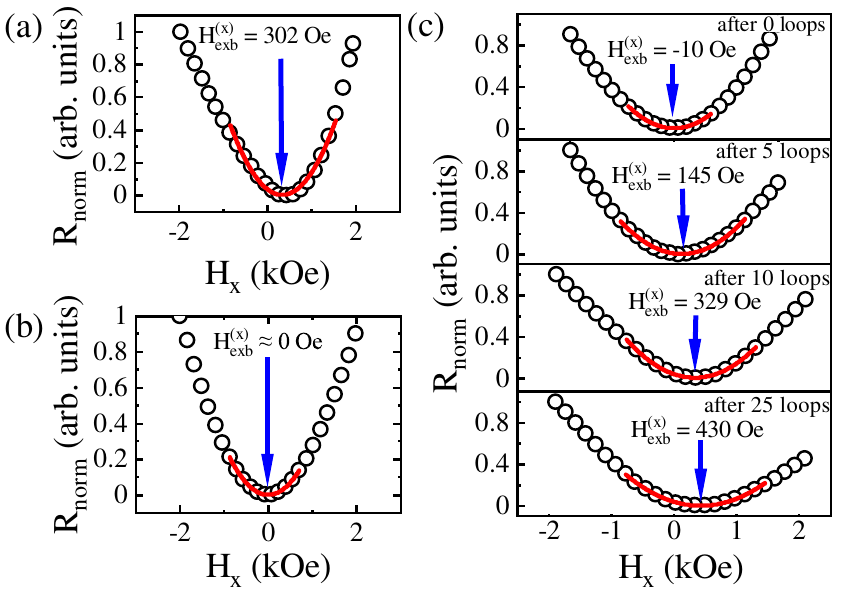}
	\caption{Normalized longitudinal magnetoresistance of Pt(5.1)/Co(0.7)/NiO-based device: (a) before the current switching experiment, (b) after current switching without an external magnetic field, (c) measured after specified number of switching cycles. The red line corresponds to a quadratic function fitted to the data in order to determine the shift.}
	\label{fig:stab_exb}
\end{figure}

Furthermore, for the Pt-based system we investigated how the number of switches affects $H_{\text{exb}}^{(x)}$.
For this purpose, we measured longitudinal magnetoresistance signal, $R_{\text{xx}}$, as in Sec.IV B, before current switching experiment, and we found that $H_{\text{exb}}^{(x)}$ was 302 Oe (Fig.~\ref{fig:stab_exb}(a)).
First, the sample was switched several times in zero external magnetic field. It was found that $H_{\text{exb}}^{(x)}$ decreases to 0 Oe (Fig.~\ref{fig:stab_exb}(b)). This proves that thermal energy generated during the pulses leads to degradation of the $H_{\text{exb}}^{(x)}$ component. The same $H_{\text{exb}}^{(x)}$ reduction effect at switching without external magnetic field was noticed by Razavi et al.~\cite{razavi_joule_2017}.  Then, by applying an in-plane $H_{\text{x}}$ field of -200 Oe, the multiple switching events were repeated. It turns out, that as the number of switching cycles increases, the $H_{\text{exb}}^{(x)}$ rises up to 430 Oe after 25 switching loops as depicted in Fig.~\ref{fig:stab_exb}(c).
\begin{figure}[h!]
	\centering
	\includegraphics[width=1\linewidth]{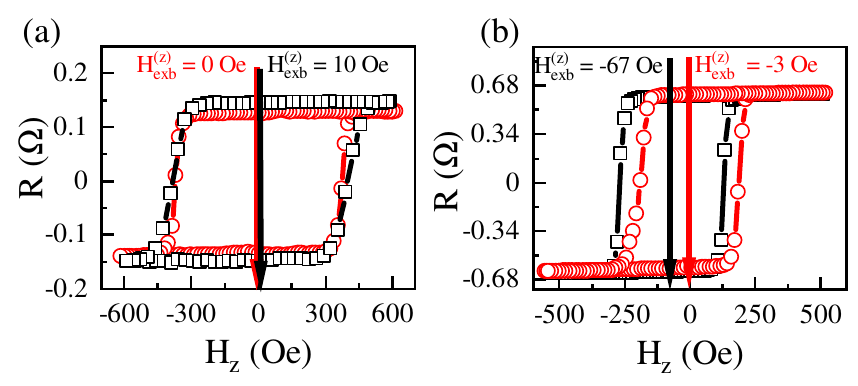}
	\caption{AHE loops before switching experiments (black points) and after multiple switching cycles (red points) Pt(5.1)/Co(0.7)/NiO (a) and annealed W(4.3)/Co(0.7)/NiO device (b) system, respectively.}
	\label{fig:stab_ahe}
\end{figure}
Finally, the AHE loop shapes and the $H_{\text{exb}}^{(z)}$ in both systems were analyzed by comparing the loops before and after the current switching experiments. No significant degradation of PMA was found, however $H_{\text{exb}}^{(z)}$ was reduced in both cases, as shown in Fig.~\ref{fig:stab_ahe}, to 0 Oe and -3 Oe, respectively for Pt and W-based devices. We conclude that during the switching events, a significant Joule heating is generated, which may lead to the temperature increase above the N\'eel temperature. If the in-plane magnetic field is applied, an increase of in-plane exchange bias is accompanied by decrease of the perpendicular exchange bias component.

\section{Summary}
\label{sec:summary}
In summary, the SOT-induced magnetization switching of (Pt,W)/Co/NiO with perpendicularly magnetized Co layer and varying HM thickness was examined. Both the in-plane $H_{\text{exb}}^{(x)}$ and perpendicular $H_{\text{exb}}^{(z)}$ exchange-bias were determined with current-driven switching, magnetoresistance and AHE methods. We demonstrate in both Pt/Co/NiO and W/Co/NiO systems the deterministic Co magnetization switching without external magnetic field which was replaced by in-plane exchange bias field.
For several selected Hall-bar devices in both systems, threshold current densities were analyzed based on our theoretical model allowing us to estimate effective parameters $\theta_{\text{SH,eff}}$ and $H_{\text{K,eff}}$. Due to higher $\theta_{\text{SH,eff}}$ in W- than Pt-based system, approximately one order of magnitude smaller critical switching current density was found.
The switching stability experiments confirm the ability to induce $H_{\text{exb}}^{(x)}$ by thermal effects. Finally, we showed wide range of resistance changes in field-free magnetization switching in case of the (Pt, W)/Co/NiO system.

\section*{Acknowledgments}
This work was supported by the National Science Centre Poland Grant No. UMO-2016/23/B/ST3/01430 (SPINORBITRONICS). WS acknowledges National Science Centre Grant Poland Grant No. 2015/17/D/ST3/00500. JK and WP acknowledge National Science Centre Poland Grant No. 2012/05/E/ST7/00240. PK and MK acknowledge the National Science Centre Poland Grant No. UMO-2015/18/E/ST3/00557. The authors would like to express their gratitude to dr. A. Kozioł-Rachwał and dr. M. Ślęzak for fruitful discussions and technical assistance in the XAS measurements. Nanofabrication was performed at the Academic Center for Materials and Nanotechnology of AGH University of Science and Technology. XAS experiment was performed at the SOLARIS National Synchrotron Radiation Centre.

\bibliography{current_switching}

\end{document}